# The perception of Architectural Smells in industrial practice

Darius Sas, Ilaria Pigazzini, Paris Avgeriou, Francesca Arcelli Fontana

## Abstract

Architectural Technical Debt (ATD) is considered as the most significant type of TD in industrial practice. In this study, we interview 21 software engineers and architects to investigate a specific type of ATD, namely architectural smells (AS). Our goal is to understand the phenomenon of AS better and support practitioners to better manage it and researchers to offer relevant support. The findings of this study provide insights on how practitioners perceive AS and how they introduce them, the maintenance and evolution issues they experienced and associated to the presence of AS, and what practices and tools they adopt to manage AS.

## Introduction

The metaphor of Technical Debt (TD) reflects the technical compromises that software practitioners make in order to achieve a *short-term* advantage at the expense of creating a technical context that increases complexity and cost in the *long-term* [Avgeriou].

Technical debt can be incurred throughout the entire software development process, so multiple types of TD can be identified (e.g. requirements, architecture, code) [Alves]. Architectural technical debt (ATD) was found to be one of the most significant types of TD, as, typically, key architectural decisions are made very early in the software lifecycle and thus have a stronger impact [Ernst].

One type of ATD are architectural smells (AS): *all AS instances are ATD items but not all ATD items are AS* [Verdecchia]. AS are defined as *"commonly (although not always intentionally) used architectural decisions that negatively impact system quality"* [Garcia]. AS manifest themselves in the system as undesired dependencies, unbalanced distribution of responsibilities, excessive coupling between components as well as in many other forms that break one or more software design principles and good practices, ultimately affecting maintainability and evolvability [Lippert]. We note that the presence of AS does not always inevitably indicate that there is a problem, but it points to places in the system's architecture that should be further analysed [Lippert].

Despite the recent attention from the research community on the topic [Verdecchia], few studies investigated how practitioners understand AS and experience the associated maintainability issues in the real world [Arcelli]. To address this shortcoming, we interviewed 21 software developers and architects to collect their opinions and experiences from industrial practice. Specifically, we focus on how practitioners *perceive* AS, what *maintenance and evolution issues* they associate with AS, how they *introduce* them and how they *deal* with them in terms of adopted practices and tools. The goal is to enrich the understanding of researchers on AS and inform practitioners on how AS manifest themselves in a real-world scenario, ultimately supporting better AS management.

While there exist several types of AS in the literature, we limited our scope to the four types of AS that are detected by most of the available tools [Azadi] and that are among the most important types of AS currently described in the literature [Arcelli]:

- **Cyclic Dependency** (CD) is defined as a set of software artefacts (e.g. classes, files, packages, components, etc.) that depend upon each other, thus creating a circle. CD breaks the Acyclic Dependencies Principle [Martin] and increases coupling.

- **Hublike Dependency** (HL) is defined as an artifact that has an excessive number of incoming and outgoing dependencies, thus creating a hub. HL breaks the modularity of the system as the hub is overloaded with responsibilities and exacerbates the dependency structure of the system.

- **Unstable Dependency** (UD) is defined as a package (or any similar construct - e.g. a component) that has too many dependencies to packages that are less stable than itself, thus increasing its reasons to change. A package is said to be *stable* if it is resilient to changes in neighbouring packages. UD breaks the Stable Dependency Principle ("*Depend in the direction of stability*") [Martin] because the affected package depends on packages less stable than itself.

- **God Component** (GC) is defined as a package (or component) whose size (measured using LOC) is noticeably bigger than the other components in the system [Lippert]. GC breaks system modularity and aggregates too many concerns into a single package.

It is important to note that the participants in our study were asked not to limit themselves to these four smells only and were free to mention experiences related to different types of AS.

## Study Design [SIDEBAR]

We performed a case study to collect experiences from industry regarding three research questions:

> **RQ1**. *How are AS perceived by practitioners?*
>
> **RQ2**. *What are the maintainability and evolvability issues experienced by practitioners that relate to the presence of AS in the system?*

*RQ3. How do practitioners introduce and deal with AS?*

For practitioners, answering these questions can help them understand and relate to issues experienced by other practitioners, obtain deeper knowledge about AS, and learn about how to manage them. Researchers, on the other hand, can understand better the real-world problems experienced by practitioners and how AS contribute to TD exactly.

We collected data by interviewing 21 practitioners from 3 companies in Europe operating in two different domains (Embedded Systems and Enterprise Applications Development) with three main programming languages (C, C++ and Java). The first company provided 12 participants, the second 6 and the third 3. The practitioners' background varies from a few years of activity (junior developers) up to 25 years of practice (architects). The average size of their projects is about 50 Million Lines of Code (LOC) for the first company, from 500.000 to 1.000.000 LOC for the second and from 250.000 to 750.000 LOC for the third. Interviews were semi-structured and each lasted approximately 30 minutes. We chose to use interviews because they allow for follow-up questions and clarifications, ensuring that participants have understood the questions. Further details about the design of this study can be found in the online appendix [Appendix].

# Results

## How AS are perceived (RQ1)

Participants reported being the most familiar with GC among the four studied AS; several practitioners reported personal experiences in managing this kind of smell. GC is perceived as a common cause of maintenance issues as well as reduced evolvability of the affected component, mainly as a result of the high level of complexity that characterizes its instances. In particular, almost all practitioners, except for two architects, had rather strong opinions on this AS and underlined its importance vividly. The two architects, instead, expressed some skepticism when discussing its importance and disregarded it as they saw no added technical value in splitting a GC.

Opinions on CD were generally aligned, and most participants considered CD as detrimental for maintainability, reliability, and testability. Concerns about reliability (e.g. deadlocks) were mostly expressed by the participants working on C/C++ projects, highlighting that even if some CD instances have not caused issues yet, they pose a high risk for future undertakings. On the other hand, participants working with Java perceived it as less detrimental than other smell types like GC. This difference in perception is probably due to the different application domains of the companies, and not only because of the differences between Java and C/C++.
We note that, typically, architectural smells are the symptom of a bigger, and more profound, issue in the architecture [Lippert] that needs to be studied case-by-case. However, in cases where CD affected reliability and testability, their very presence was considered as the problem that developers were trying to resolve.

Opinions were much more polarized when the HL smell was discussed. Some participants mentioned that: (1) it should not be considered a problem because it could be a result of an intentional design decision; (2) it should not be a cause of concern as long as it is understandable; and, (3) as one participant expressed, it is easy to solve. However, other participants (and especially the ones working with Java) mentioned that HL is very important to avoid because it is not easy to manage and it hinders both maintainability and evolvability by making it hard to understand how to insert new code in the presence of a HL.

Concerning UD, participants generally perceived it as a threat to both maintainability and evolvability, highlighting their concerns about the change ripple effects associated with UD and underlining the importance of *avoiding dependencies towards packages that constantly evolve*. Nevertheless, one developer expressed their doubts about the importance of this AS while few more outlined that they did not fully understand it and gave no feedback about it.

From these results, it appears that, while all AS are considered detrimental, they are perceived differently by practitioners depending on their past experiences, educational background, and application domain: GC and CD are perceived as the most important ones, HL is considered "*manageable*", and UD is considered detrimental but not critical. It is important also to take into account that UD is less visible than the other smells: one cannot tell by looking at a package that it is less stable than another one without employing dedicated tooling.

Finally, we observed the existence of a slight correlation between the experience of interviewees and the type of concerns expressed about an AS. Junior participants tended to be more concerned about the short-term problems (e.g. presence of CDs and impact deployed system), while senior participants were keener on long-term evolvability and team-related matters (e.g. new team members making changes to a GC).

## How AS impact maintenance and evolution (RQ2)

The participants discussed plenty of anecdotes and experiences about maintenance and evolution issues that they associated with the presence of AS. Almost all anecdotes about GC involve the difficulty of understanding the functionality provided by the component, mainly caused by the excessive internal entanglement of files (or classes), the excessive amount of functionality implemented, and the way functionality is scattered across the component. The relationship between GC and code duplications was also frequently discussed. Components affected by GC do not provide fine-grained classes that can be easily reused inside or outside the component, but large and entangled classes. Hence, when developers need to reuse an existing functionality, they prefer to copy the entire class and adapt it for the new purpose, instead of extracting a small, reusable functionality. On top of creating duplicated code, this also further enlarges the existing GC.

The experiences about CD are rather diverse and range from dealing with deadlocks and low throughput to unclear chain of command between components and poor separation of concerns in general. Cycles were also reported as an "*intertwined mess*" that is hard to understand; e.g. when there is a package that requests data from another package which in turn requests it back from the initial package. These problems resulted in a significant amount of effort required to be fixed or dealt with along the way, and in some cases, they showed up only in production or at the customer. Participants also mentioned problems that had a more widespread impact; for example, a cycle prevented the creation of a microservice out of a subset of packages, as all the packages in the cycle had to be included in the microservice (the desired functionality could not be isolated).

Concerning HL, practitioners associated it with two types of issues: (1) difficulty of understanding the logic in the central component and (2) change ripple effects propagating from the components that the central component depends upon to the components depending on it, mentioning also a possible overlap with UD. The former was usually associated with how the central component exposes its functionality through its interface. The latter caused changes to unexpected parts of the system that practitioners did not expect to relate to the initial change, during activities such as bug fixing.

The maintenance issues that associated with UD the most, were change ripple effects. In several instances, practitioners reported that functional changes to a certain component (or package) also required several files in other components to change as well. As reported by two participants, the possibility of changes propagating to other components increases the difficulty of making changes: practitioners are forced to only make changes compatible with the other components in order to avoid changing and recompiling those other components.

## How AS are introduced and managed (RQ3)

Participants reported their experiences in how they get to *introduce an AS* in the system. Some participants admitted that it often happens by design; for instance, concerning GC, the component or the file *is intended* to be large. Subsequently, as reported by other interviewees, developers tend to underestimate the severity of the introduced GC, while the incremental changes applied to it contribute in making it even larger.

In other cases, AS are introduced inadvertently. For example, the participants reported that a bad separation of concerns at design time or the wrong exploitation of class inheritance can result in CD. Another participant mentioned that they used to create a dedicated interface to hide unstable components behind it as a "practice" to avoid the propagation of changes; however, this is precisely the description of a UD smell, being misinterpreted as a good practice.

In many cases, introducing AS seems unavoidable and accepted as a "*necessary evil*". For example, one participant explained that in view of an imminent deadline, they focus on developing the new feature and having a first structure of the code, without caring about its maintainability.

Moving on to the *management of AS*, we asked the participants about their experiences with *AS refactoring*. Most of them had experience with the refactoring of GC, in particular the practice of splitting the component in smaller pieces by applying incremental changes or by detaching the smallest, easiest sub-components first. One interviewee managed to break a CD by re-modelling the involved dependencies to follow a hierarchical structure; others reported creating replacement interfaces and slowly migrating clients to them while refactoring the existing components. In contrast, developers do not commonly refactor HL because of the required effort; if they can, they tend to "*code around it*" without removing it when developing new features, allowing it to persist. One interesting reason mentioned for *not* refactoring AS is the absence of a comprehensive regression test suite.

Concerning practices which support the refactoring of AS, some participants mentioned the usage of SonarQube to keep the code readable and maintainable; this can ease the refactoring of AS since often the poor quality of the code makes refactoring even more difficult and time-consuming. Another indicated pair programming and the help of senior developers as valid support. However, not all the interviewees reported the adoption of refactoring practices. Some even pointed out that they avoid refactoring because their clients do not pay for refactoring time and as long as the system has no visible problems in production, they do not intervene.

Finally, we also asked whether practitioners *use tools* to manage architectural smells. SonarQube was mentioned by quite a few participants, but only once in regard to an AS (i.e. to detect cycles). Besides that, practitioners do not rely on any specific tool to manage AS. Nonetheless, participants did mention ideal features that they would like to have in an ideal tool that manages AS. Due to space limitations, the features are reported in our online appendix [Appendix], and we created a mind map to summarise the results of all three RQs in Figure 1.

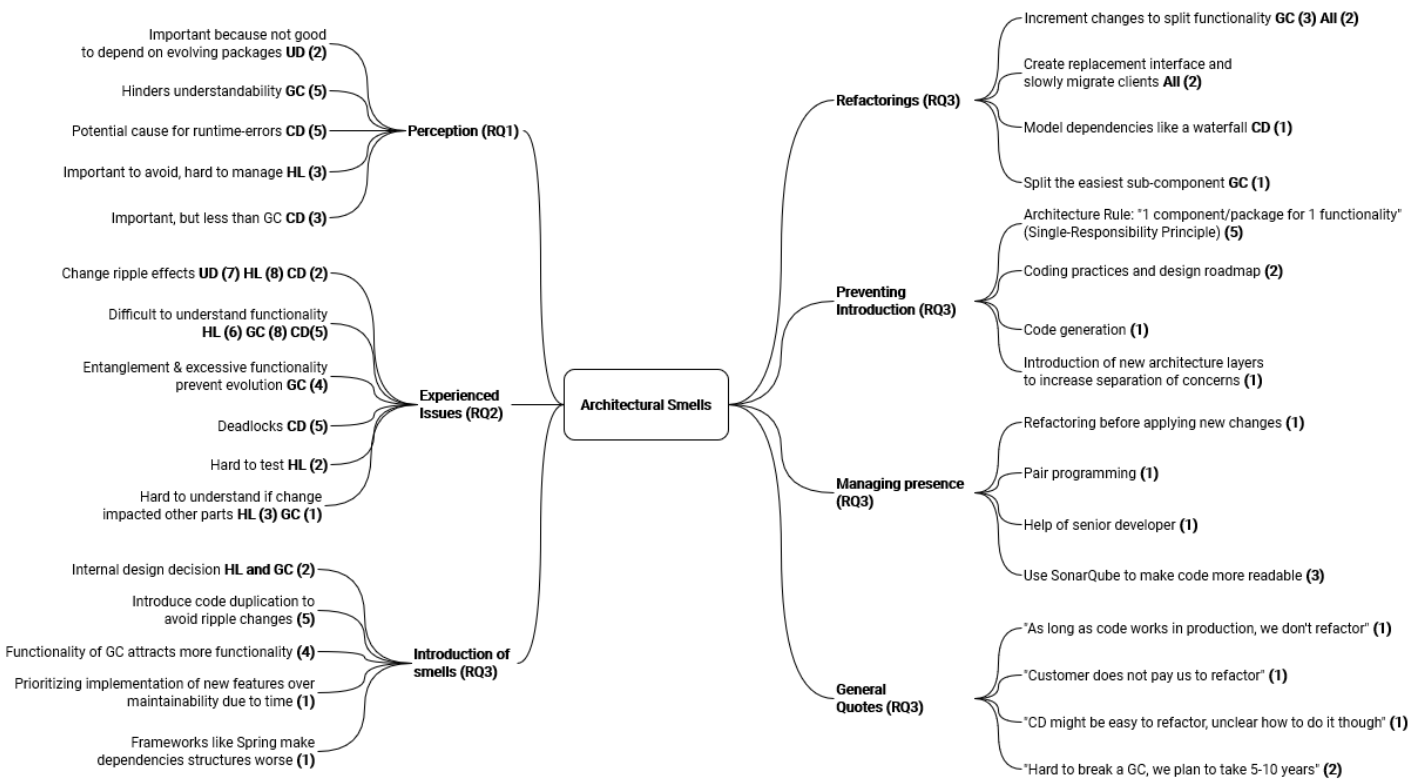

**Figure 1.** Mind map summarising the perception, experiences, prevention, introduction, and presence of architectural smells as described by the participants. In parentheses we report the number of data points, and, if appropriate, the type of associated AS.

## Discussion and implications

The presented results indicate that AS clearly help incurring ATD: they have a direct, architecture-level impact on the maintainability and evolvability of the affected parts. AS make changes harder to implement by increasing the effort required to understand the implications of a change, making it *easy* to *underestimate* the effort necessary for the change, and *hard* to *plan* ahead. Practitioners are aware and well-informed about good design practices, but they struggle following them diligently, often prioritizing delivering a feature over good design. Fowler calls this *reckless* and *deliberate* TD [Fowler], because practitioners understand the long-term implications of their decisions but still decide to incur technical debt. By doing so, practitioners are forced, sooner rather than later, to apply refactorings before proceeding with the implementation of new features (as mentioned by the participants) and pay a considerable amount of TD interest every time they need to extend the system.

As emerged from the interviews, TD is also incurred *inadvertently* [Fowler], either *recklessly*, because of poor knowledge about the design of the parts affected by change (e.g. a component requesting a parameter that belongs to itself from another component), or *prudently*, because the optimal design solution only becomes clear after implementing the chosen solution. The introduction of technical debt through non-optimal solutions that is then detected as AS is not automatically controlled, as we observed a lack of adoption of tooling dedicated to manage AS - practitioners mostly focus on code TD.

At any rate, regardless of the how, incurring TD is inevitable and inherent to the software development process, so practitioners must adopt practices that enable its management. Similarly to any other type of TD items, the first step in managing AS is *detecting* them. Azadi et al. provide a recent list of tools that detect AS [Azadi] for practitioners to consider. Another, even more important step is prevention. Practitioners should pay particular attention to how they create internal dependencies as there is a fine balance between Changeability and number of dependencies per file: too many, and files become entangled, making the system hard to modify and giving rise to GC and CD; too few, and the system is also hard to modify, because fewer classes are reused (tree-like dependency graph [Lippert]) resulting in multiple classes implementing similar functionality and applying the same change to all of them is repetitive. Therefore, practitioners should carefully balance how these dependencies are created by devising clear architectural rules that prevent the creation of undesired dependencies that end up generating AS.

Our research work to date has focused precisely on addressing these issues, culminating in the development of Arcan, a tool to make AS detection and dependency analysis as easy as possible to practitioners that work either in Java or C/C++, and soon Python and C#.

# Biographies

**Darius Sas**

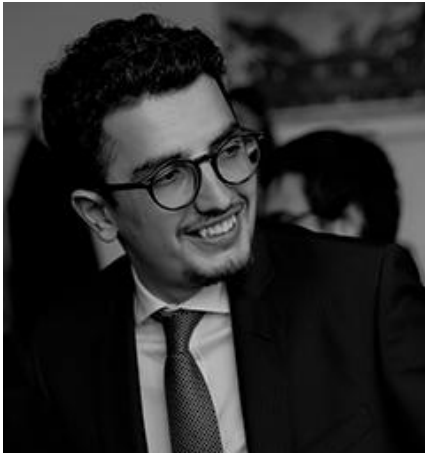

Darius Sas is a Ph.D. student at the Bernoulli Institute for Mathematics, Computer Science and Artificial Intelligence, University of Groningen, the Netherlands. He received his Master's degree in Computer Science in 2018 from the University of Milano-Bicocca, Milan, Italy. He then started his academic career as a Ph.D. student in the Software Engineering and Architecture (SEARCH) group led by Paris Avgeriou at the University of Groningen. His Ph.D. project focuses on architectural technical debt elimination in embedded systems and is part of a European project (SDK4ED) that focuses on the interplay between technical debt, energy efficiency, and dependability. Contact him at d.d.sas@rug.nl.

**Ilaria Pigazzini**

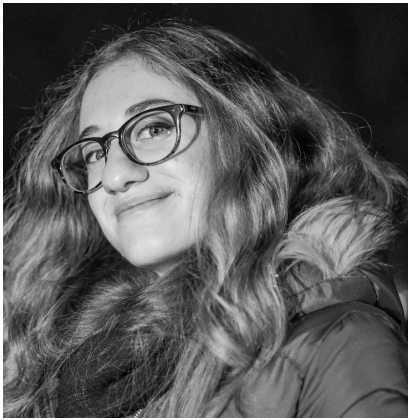

Ilaria Pigazzini is currently a Ph.D. student in computer science at the Department of Computer Science, Systems and Communications, University of Milano-Bicocca. She has received her B.Sc. and M.Sc. degrees from the University of Milano-Bicocca in Computer Science in 2016 and 2018, respectively. Her research interests include reverse engineering, architectural smell detection and refactoring of Object Oriented systems. Contact her at i.pigazzini@campus.unimib.it.

**Paris Avgeriou**

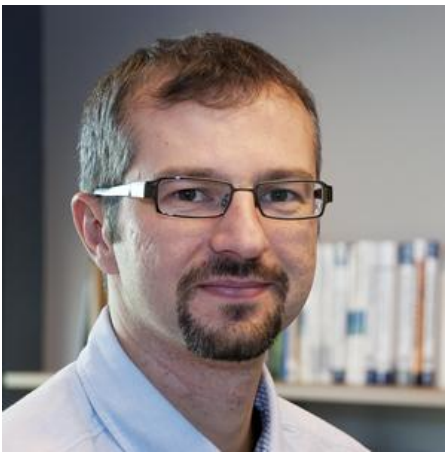

Dr. Paris Avgeriou is Professor of Software Engineering at the University of Groningen, the Netherlands where he has led the Software Engineering research group since September 2006. Before joining Groningen, he was a post-doctoral Fellow of the European Research Consortium for Informatics and Mathematics. He is the Editor in Chief of the Journal of Systems and Software, as well as an Associate Editor for IEEE Software. He also sits on the board of the Dutch National Association for Software Engineering (VERSEN) and the Dutch research school IPA. He has co-organized several international conferences such as ECSA, ICSA, and Tech Debt and served on their steering committees. His research interests lie in the area of software architecture, with strong emphasis on architecture modeling, knowledge, evolution, analytics and technical debt. He champions the evidence-based paradigm in Software Engineering research and works towards closing the gap between industry and academia. Contact him at p.avgeriou@rug.nl.

**Francesca Arcelli Fontana**

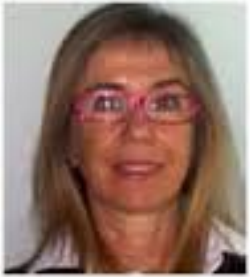

Francesca Arcelli Fontana had her Master degree and Ph.D. in Computer Science taken at the University of Milano. She is currently in the position of Full Professor at University of Milano Bicocca. The actual research activity principally concerns the software engineering field. In particular in software evolution, reverse engineering, managing technical debt, and software quality assessment. She is at the head of the Software Evolution and Reverse Engineering Lab at University of Milano Bicocca. Contact her at francesca.arcelli@unimib.it.